\documentclass[aps,prl,reprint,nofootinbib]{revtex4-2}
\usepackage{enumerate}

\usepackage{amsthm}
\usepackage{amssymb}
\usepackage{amsmath}

\newtheorem{example}{Example}

\usepackage{soul}
\usepackage{ulem}
\usepackage{cancel}
\usepackage{color}
\usepackage{xcolor}

\begin{document}
	
	
	\title{Equal opportunities lead to maximum wealth inequality}
	
	\author{Ben-Hur Francisco Cardoso}
	\email{ben.cardoso@posgrad.ufsc.br}
	\affiliation{Departamento de Economia e Relações Internacionais, Universidade Federal de Santa Catarina, Florianópolis, SC, Brazil}
	\author{Sebastián Gonçalves}
	\email{sgonc@if.ufrgs.br}
	\affiliation{Instituto de Física, Universidade Federal do Rio Grande
		do Sul, Porto Alegre, RS, Brazil}
	\author{José Roberto Iglesias}
	\email{iglesias@if.ufrgs.br}
	\affiliation{Instituto de Física, Universidade Federal do Rio Grande
		do Sul, Porto Alegre, RS, Brazil}
	\affiliation{Instituto Nacional de Ciência e Tecnologia de Sistemas
		Complexos, INCT-SC, CBPF, Rio de Janeiro, RJ, Brazil}

	\date{\today}
	
	\begin{abstract}
		\noindent  If wealthier people have advantages in having higher returns than poor, inequality will unequivocally increase, but is equal opportunity enough to prevent it? According to several models in economics and econophysics, no. They all display wealth concentration as a peculiar feature of its dynamics, even though no individual can have repeating gains advantage. Here, generalizing these particular models, we present a rigorous analytical demonstration, using master equation formalism, that any fair market that gives each agent the same expected return conducts the system to maximum inequality.
	\end{abstract}
	
	\maketitle
	
	The remarkable growth of wealth concentration during the XXIth Century~\cite{chancel2019ten} in almost all countries is often attributed to some rich-biased advantage~\cite{diprete2006cumulative} due to market failures, producing a self-feeding cascade~\cite{moukarzel2011multiplicative}.
	Although this type of mechanism promotes wealth concentration indeed, we could be tempted to think that it would not happen in a hypothetical efficient or fair market, where no individual has repeating advantages~\cite{fargione2011entrepreneurs,biondi2019inequality,biondi2020financial,klass2007forbes,levy2003investment}.
	
    Let us start by reviewing some models which show us that such reasoning may be not correct.
	Consider a system of $N$ agents, where $x_{i,t} \geq 0$ is the wealth of agent $i$ at time $t$. Since agents face risk, the future market opportunities are not predicable and the realized growth rate of individual wealth of a given agent in the transition $t\rightarrow t+1$ can be greater or lower than the average. However, an  efficient market prevents the arbitrage opportunities, {\it i.e.}, the expected growth rate of individual wealth is the same for all agentes. Therefore, there exists a positive constant $\alpha_t$ such that~\cite{fargione2011entrepreneurs,biondi2019inequality,biondi2020financial,klass2007forbes,levy2003investment,hayes2002,iglesias2012entropy,bouleau2017impact,cardoso2020wealth,gusman2005wealth, bustos2012yard, laguna2005economic}
	\begin{equation}\label{eq:fair_intro}
		\mathbb{E}_t[x_{i,t+1}] = \alpha_t\>x_{i,t}
	\end{equation}
	for all agent $i$. In other words, no agent is systematically favored. That is why we can call this market fair, with equal opportunities. However, all studied models satisfying Eq.~\ref{eq:fair_intro} --random growth models~\cite{fargione2011entrepreneurs,biondi2019inequality,biondi2020financial,klass2007forbes,levy2003investment} and unbiased kinetic exchange models~\cite{hayes2002,iglesias2012entropy,bouleau2017impact,cardoso2020wealth,gusman2005wealth, laguna2005economic,cardoso2021wealth}-- converge to the {\it condensed} state, where one agent concentrates all the disposable wealth.
	
	Within the random growth model framework, the individual wealth dynamics follows~\cite{fargione2011entrepreneurs,biondi2019inequality,biondi2020financial,klass2007forbes,levy2003investment}
	\begin{equation}\label{eq:rgm}
		x_{i,t+1} = \alpha_{i,t}\,x_{i,t}\>\>\>,\>\>\>\mathbb{E}_t[\alpha_{i,t}] = \alpha_t
	\end{equation}
	that is, at time $t$ each agent $i$ extracts a random return $\alpha_{i,t} > 0$ from a common probability density function $g_t(\alpha)$ with mean $\alpha_t$~\footnote{In these models, it is generally assumed that $\alpha_t > 1$, that is, the average wealth grows with time.}. 
	
	In the unbiased kinetic exchange models, two agents are sequential and randomly chosen to exchange wealth. So, let assume a wealth exchange between random selected agents $i$ and $j$
	\begin{equation}\label{eq:kem}
		x_{i,t+1} = x_{i,t} + \Delta_{i,t} \>\>\> \text{and} \>\>\> x_{j,t+1} = x_{j,t} + \Delta_{j,t},
	\end{equation}
	where $\Delta_{i,t}$ is the stochastic gain of agent $i$ such that $\Delta_{i,t} + \Delta_{j,t} = 0$. In mathematical terms, this zero-sum exchange process describe an efficient market if $\mathbb{E}_t[\Delta_{i,t}] = \mathbb{E}_t[\Delta_{j,t}] = 0$~\cite{bouleau2017impact,cardoso2021wealth}.
	
	The best-known unbiased kinetic exchange model is the {\it Yard-Sale:}~\cite{hayes2002} 
	\begin{equation}\label{eq:yard}
		\Delta_{i,t} = \eta_t\lambda_t\min(x_{i,t},x_{j,t}),\>\>\eta_t\in\{-1,1\},\>\>\mathbb{E}_t[\eta_t] = 0,
	\end{equation}
	where $0\leq\lambda_t\leq 1$ can be a constant or a random number. This rule was generalized in two ways. First, restricting exchanges between agents linked by a static~\cite{bustos2012yard} or dynamic~\cite{gusman2005wealth, laguna2005economic} network. Second, introducing agent specific parameters $\lambda_{i}$~\cite{caon2007unfair, cardoso2020wealth} such that
	\begin{eqnarray}\label{eq:min}
		\nonumber
		\Delta_{i,t} = \eta_t\min\Big(\lambda_ix_{i,t},\lambda_jx_{j,t}\Big),\>\>\eta_t\in\{-1,1\},\\
		\>\>\mathbb{E}_t[\eta_t] = 0,
	\end{eqnarray}
	where $\lambda_{i}$ is the fraction of the wealth that agent $i$ will risk in the exchange. 
	
	Also, it was proposed an exchange rule such that~\cite{bouleau2017impact}
	\begin{eqnarray}\label{eq:loser}
		\nonumber
		\Delta_{i,t} = \epsilon_t\lambda_t x_{j,t} - (1 - \epsilon_t)\lambda_t x_{i,t},\>\>\epsilon_t\in\{0,1\},\>\>\\
		\mathbb{E}_t[\epsilon_t] = \frac{x_{i,t}}{x_{i,t} + x_{j,t}},
	\end{eqnarray}
	where $0 \leq \lambda_t \leq 1$ can be a random or constant number. Finally, a similar model was proposed by one of the authors~\cite{iglesias2012entropy}, where
	\begin{equation}\label{eq:iglesias}
		\Delta_{i,t} = \eta_t\frac{x_{i,t}x_{j,t}}{x_{i,t} + x_{j,t}},\>\>\eta_t\in\{-1,1\},\>\>\mathbb{E}_t[\eta_t] = 0.
	\end{equation}
	
	As stated above, all these models (Eqs.~\ref{eq:rgm},~\ref{eq:yard},~\ref{eq:min},~\ref{eq:loser} and~\ref{eq:iglesias}) converge to the condensed state. In a previous paper~\cite{cardoso2021wealth}, working with a master equation in the thermodynamic limit ($N \rightarrow \infty$), we proved that condensation is the asymptotic state of any model with agents only characterized by their wealth, whose dynamics are given by binary exchanges and without wealth growth (the unbiased kinetic exchange model). Here, we extend the result by relaxing these three assumption, proving that any  efficient market satisfying Eq.~\ref{eq:fair_intro} converges to maximum inequality. 
	
	\subsection{The master equation}\label{sec:master}
	Let be a system of agents where each one is in a state from the set of all possible ones represented by $\Omega $. For notation simplicity, if an agent is in the state $\Gamma = (\Gamma^0,\Gamma^1,\cdots) \in \Omega$, $\Gamma^0 = x$ is the agent's wealth. The system can be represented by the probability density function $p_t(\Gamma)$, where $p_t(\Gamma)d\Gamma$ is the fraction of agents within $d\Gamma$ of $\Gamma$ at time $t$.  This probability density function must obey the following properties:
	\begin{enumerate}[(i)]
		\item Negative wealth is not allowed, then, for all $t$
		\begin{equation}
			x < 0 \Rightarrow p_t(\Gamma) = 0.
		\end{equation}
		\item For all $t$, the probability density function is normalized
		\begin{equation}\label{eq:norm_x}
			\int_{\Omega}d\Gamma\>p_t(\Gamma) = 1.
		\end{equation}
		\item For all $t$, we define the average wealth as
		\begin{equation}
			\label{eq:first_x}
			\mu_{t} \equiv \int_{\Omega}d\Gamma\>x\>p_t(\Gamma).
		\end{equation}
	\end{enumerate}
	
	The dynamics of $p_t(\Gamma)$ can be represented by the Master Equation~\cite{katriel2015directed,pitaevskii2012physical} 
	\begin{eqnarray}\label{eq:master}
		p_{t+1}(\Gamma) = \int_{\Omega}d\Gamma'\>p_t(\Gamma')\>\omega_t(\Gamma'\rightarrow \Gamma,p_t)
	\end{eqnarray}
	where $\omega_t(\Gamma'\rightarrow \Gamma,p_t)\>d\Gamma$ is the probability of an agent to be within $d\Gamma$ of $\Gamma$ at time $t+1$, conditioned on being in the state $\Gamma'$ at time $t$~\cite{pitaevskii2012physical}. For simplicity, we use the notation $\omega_t(\Gamma'\rightarrow\Gamma) \equiv \omega_t(\Gamma'\rightarrow\Gamma,p_t)$. 
	
	The transition rate $\omega_t(\Gamma'\rightarrow \Gamma)$ must obey the following properties: 
	
	\begin{enumerate}[(i)]
		\item Negative wealth is not allowed, then, for all $t$
		\begin{equation}
			\label{eq:pos}
			x < 0\>\>\text{or}\>\>x'< 0 \Rightarrow \omega_t(\Gamma'\rightarrow \Gamma) = 0.
		\end{equation}
		\item Agents live forever; that is, one agent in a given state $\Gamma' \in \Omega$ at $t$ must be in another state in $\Omega$ at  $t+1$.
		 This requires that the integral over all the  transitions must be $1$, so  we have for all $t$ that
		\begin{equation}
			\int_{\Omega} d\Gamma \>\omega_t(\Gamma'\rightarrow\Gamma) = 1\>\>\>,\>\>\>\forall\>\Gamma' \in \Omega.
			\label{eq:normal_omega}
		\end{equation}
		Also, this condition guarantees the normality of $p_t$ for all $t$ (Eq.~\ref{eq:norm_x}).
		\item As the market is fair, the expected individual growth rate of wealth is equal for all agents, that is, there is a positive constant $\alpha_t$ for all $t$ such that
		\begin{equation}
			\label{eq:fair}
			\int_{\Omega}\>d\Gamma\>x\>\omega_t(\Gamma'\rightarrow \Gamma) = \alpha_t x'\>\>\>,\>\>\>\forall\>\Gamma' \in \Omega.
		\end{equation}
		
		\item All agents face risk, that is, even if the random growth rate has the same expected value $\alpha_t$ for all, this randomness can make the individual growth in a given instance less or greater than that average.
		So, for all $t$ and $\Gamma' \in \Omega$ such that $x' > 0$
		\begin{eqnarray}
			\label{eq:risk}
			\int_{\Omega}\>d\Gamma\>|x-\alpha_t x'|\>\omega_t(\Gamma'\rightarrow \Gamma) > 0.
		\end{eqnarray}
		This rule cannot be valid for agents with zero wealth ($x' = 0$), since it will contradict Eq.~\ref{eq:fair}.
	\end{enumerate}
	
	Using Eqs.~\ref{eq:first_x},~\ref{eq:master} and~\ref{eq:fair}, it is ease to verify that 
	\begin{equation}\label{eq:alpha}
		\mu_{t+1} = \alpha_t\mu_t.
	\end{equation}
	
	That state transition rate can be driven by an autonomous change (like random growth models) or by interactions among agents (like unbiased kinetic exchange models or any kind of exchange model among $m > 2$ agents). Below we describe two examples.
	
	\begin{example}
		The agent's state in the class of random growths models (Eq.~\ref{eq:rgm}) is restricted to wealth $\Gamma = \{x\}$. So, in terms of the master equation formulation, their dynamics can be written as
		\begin{equation}
			\omega_t(\Gamma'\rightarrow\Gamma) = \int_0^{\infty}d\alpha\>g_t(\alpha)\>\delta(\alpha\>x' - x),
		\end{equation}
		where $g_t(\alpha)$ is the probability density function of random returns at time $t$.
	\end{example}

	\begin{example}
		The agent's state in the almost all unbiased kinetic exchange models (Eq.~\ref{eq:rgm}) is restricted to wealth and a fixed risk propensity $\Gamma = \{x,\lambda\}$. For example, the dynamics of Eq.~\ref{eq:min} can be written as
		\begin{eqnarray}
			\nonumber
			\omega_t(\Gamma'\rightarrow\Gamma) = \frac{1}{2}\delta(\lambda'-\lambda)\int_{\Omega}d\Gamma''\>p_t(\Gamma'')\times\\
			\nonumber
			\bigg[\delta\Big(x - x' +\min\big(\lambda' x',\lambda'' x''\big)\Big)+\\
			\delta\Big(x -x' - \min\big(\lambda' x',\lambda'' x''\big)\Big)\bigg].
		\end{eqnarray}
	\end{example}
	
	In addition to the other models reviewed in the Introduction, the procedure made in Examples 1 and 2 can be easily extended to include higher order interactions, {\it i.e}, with exchanges among $m > 2$ agents.
	
	\subsection{The ever-increasing inequality}\label{sec:ineq}
	
	To measure inequality, we use the Gini index~\cite{sen1997economic}:
	\begin{equation}\label{eq:gini}
		G_t = \frac{1}{2}\int_{\Omega^2} d\Gamma d\Gamma_1\> \frac{| x-x_1 |}{\mu_t} \> p_t(\Gamma) p_t(\Gamma_1).
	\end{equation}
	The Gini index varies between $0$, which corresponds to perfect equality (everyone has the same wealth) and $1$, that corresponds to maximum inequality, associated with the marginal probability density function of wealth~\cite{boghosian2017oligarchy,cardoso2021wealth}
	\begin{equation}\label{eq:oli}
		\delta (x) + \lim_{N \rightarrow \infty} \frac{\delta (x - \mu_t N)}{N}.
		\end{equation}
	This means that all wealth is in hands of an infinitesimal part of the system, with measure zero. Then, this probability density function is such that $p_t(\Gamma) = 0$ for all $x > 0$~\cite{boghosian2014kinetics, boghosian2015h, boghosian2017oligarchy,cardoso2021wealth}, also satisfying Eqs.~\ref{eq:norm_x} and~\ref{eq:first_x}.
	
	Now, we can enunciate and proof the following {\bf proposition}: {\it In an efficient market, $G_{t+1} \geq G_t$ for all $t$ and }
	\begin{equation}\nonumber
		\lim_{t\rightarrow\infty}G_t = 1.
	\end{equation}
	
{\it Proof}: Given the definition of Gini index (Eq.~\ref{eq:gini}) and the master equation (Eq.~\ref{eq:master}), we have that
	\begin{eqnarray}
				\label{eq:gini_ev}
				\nonumber
				G_{t+1} = \frac{1}{2}\int_{\Omega^2} d\Gamma d\Gamma_1 \frac{| x-x_1 |}{\mu_{t+1}}  p_{t+1}(\Gamma) p_{t+1}(\Gamma_1) =\\
				\nonumber
				\frac{1}{2}\int_{\Omega^2}d\Gamma'd\Gamma_1' p_t(\Gamma')p_t(\Gamma_1')\times\\
				\int_{\Omega^2}d\Gamma d\Gamma_1\frac{|x-x_1|}{\mu_{t+1}}\omega_t(\Gamma'\rightarrow \Gamma) \omega_t(\Gamma_1'\rightarrow \Gamma_1).
		\end{eqnarray}
		
		Now, since the integral of absolute value is always equal or grater than the absolute value of the integral~\cite{hansen2003jensen}, the Eqs.~\ref{eq:normal_omega},~\ref{eq:fair} and~\ref{eq:alpha} implies
		\begin{eqnarray}\label{eq:abs_ineq}
				\nonumber
				\int_{\Omega^2}d\Gamma d\Gamma_1\frac{|x-x_1|}{\mu_{t+1}}\omega_t(\Gamma'\rightarrow \Gamma)  \omega_t(\Gamma_1'\rightarrow \Gamma_1) \geq \\
				\frac{\alpha_t|x'-x_1'|}{\mu_{t+1}} =\frac{|x'-x_1'|}{\mu_t}.
		\end{eqnarray}
		So, given the Eqs.~\ref{eq:gini_ev},~\ref{eq:abs_ineq}, and~\ref{eq:gini}, we find that
		\begin{eqnarray}
				\nonumber
				G_{t+1} = \frac{1}{2}\int_{\Omega^2}d\Gamma'd\Gamma_1'p_t(\Gamma')p_t(\Gamma_1')\times\\
				\nonumber
				\int_{\Omega^2}d\Gamma d\Gamma_1\frac{|x-x_1|}{\mu_{t+1}}\omega_t(\Gamma'\rightarrow \Gamma) \omega_t(\Gamma_1'\rightarrow \Gamma_1)\geq\\
				\nonumber
				\frac{1}{2}\int_{\Omega^2}d\Gamma'd\Gamma_1'\frac{|x'-x_1'|}{\mu_t}p_t(\Gamma')p_t(\Gamma_1') \\
				\Rightarrow G_{t+1} \geq G_t
		\end{eqnarray}
		
		Since $G_t \leq 1$ for all $t$, we immediately find that $G_t = 1 \Rightarrow G_{t+1} = G_t$.
		Now, let us proof that reciprocate: $G_{t+1} = G_t \Rightarrow G_t = 1$. From Eqs.~\ref{eq:gini} and~\ref{eq:gini_ev}, $G_{t+1} - G_t = 0$ means that
			\begin{eqnarray}\label{eq:dG}
				\nonumber
				0 =\frac{1}{2}\int_{\Omega^2}d\Gamma' d\Gamma_1'p_t(\Gamma')p_t(\Gamma_1') \bigg[-\frac{|x'-x_1'|}{\mu_t} + \\
				\int_{\Omega^2}d\Gamma d\Gamma_1\frac{|x-x_1|}{\mu_{t+1}}\omega_t(\Gamma'\rightarrow \Gamma) \omega_t(\Gamma_1'\rightarrow \Gamma_1)\bigg].
			\end{eqnarray}
			
			By the Eq.~\ref{eq:abs_ineq}, the term in square brackets of Eq.~\ref{eq:dG} is always non-negative, so the equality only holds if 
			\begin{eqnarray}\label{eq:cond}
				\nonumber
				\int_{\Omega^2}d\Gamma d\Gamma_1\frac{|x-x_1|}{\mu_{t+1}}\omega_t(\Gamma'\rightarrow \Gamma) \omega_t(\Gamma_1'\rightarrow \Gamma_1) = \\
				\frac{|x'-x_1'|}{\mu_t}
			\end{eqnarray}	
			for all $\Gamma',\Gamma_1'\in\Omega$ such that $p_t(\Gamma') > 0$ and $p_t(\Gamma_1') > 0$. By Eq.\ref{eq:fair}, the particular case of $x = 0$ trivially satisfies the Eq.~\ref{eq:cond}. Furthermore, we must have $p_t(\Gamma') = 0$ for all $x' > 0$. Otherwise, there would be at least one $x' > 0$ such that $p_t(\Gamma') > 0$; in that case, for $x_1' = x'$, the left side of Eq.~\ref{eq:cond} becomes
			\begin{eqnarray}
				\nonumber
				\int_{\Omega^2}d\Gamma d\Gamma_1\frac{|x-x_1|}{\mu_{t+1}}\omega_t(\Gamma'\rightarrow \Gamma) \omega_t(\Gamma_1'\rightarrow \Gamma_1)\geq\\
				\nonumber
				\int_{\Omega}d\Gamma\frac{|x-\alpha_t x'|}{\mu_{t+1}}\omega_t(\Gamma'\rightarrow \Gamma) > 0,
			\end{eqnarray}
			by Eq.\ref{eq:risk}, contradicting the right side of Eq.~\ref{eq:cond} that is zero when $x' = x_1'$. So, we conclude that
			\begin{eqnarray}\label{eq:aux}
				\nonumber
				G_{t+1} - G_t = 0 \Rightarrow \Big[p_t(\Gamma) = 0\>\>\forall x > 0\Big] \Rightarrow G_t = 1.
			\end{eqnarray}
		
			Summarizing, $G_t = 1 \Leftrightarrow G_{t+1} = G_t$ and, complementary, $G_t < 1 \Leftrightarrow G_{t+1} > G_t$. Then, $\sup_{t\in \mathbb{N}}\big\{G_t\big\} = 1$. By the well known {\it monotone convergence theorem}, we find
			\begin{eqnarray}
				\lim_{t\rightarrow\infty}G_t = \sup_{t\in \mathbb{N}}\big\{G_t\big\} = 1,
			\end{eqnarray}
	that is, we asymptotically obtain condensation, Q.E.D.
	
	\subsection{Final comments}

	Previous numerical~\cite{fargione2011entrepreneurs,biondi2019inequality,biondi2020financial,klass2007forbes,levy2003investment,hayes2002,iglesias2012entropy,bouleau2017impact,cardoso2020wealth,gusman2005wealth, bustos2012yard, laguna2005economic} and analytical~\cite{biondi2020financial,bouleau2017impact,moukarzel2007wealth,boghosian2015h,cardoso2021wealth} results show that efficient markets lead to maximum inequality, or condensation, in specific model cases.
	However, our contribution is the first analytical demonstration that such fate occurs for all efficient markets, regardless of their particularities.
	
	We might think that a system without regulatory policies is {\it fair} since no individual has systematic or {\it a priori} advantages~\cite{fargione2011entrepreneurs,biondi2019inequality,biondi2020financial,klass2007forbes,levy2003investment}. However, we have demonstrated here in full generality that such apparent ``fairness'' does not prevent inequality from increasing up to a maximum.
	
	Thus, what are the prepositions' implications here demonstrated in the real world?
	 The worldwide data about economies show a clear tendency toward increasing inequalities after relaxing taxation policies ~\cite{piketty2018distributional}. 
	We can cite the US case, where the fraction of national wealth in the hands of the top $1\%$ risen from $22\%$ in $1980$ to $37\%$ in $2015$~\cite{piketty2018distributional}. Given that a ``fair'' market cannot impede inequality from increasing, some mechanism to favor the poorest is necessary, such as taxation and redistribution of wealth, to guarantee less unequal societies.
	

\end{document}